\documentclass[twocolumn,prb,aps,floatfix,superscriptaddress,amssymb]{revtex4}
\usepackage{epsfig}
\usepackage{amsmath}

\begin{document}

\title{Wigner crystallization at graphene edges}

\author{A. D. G\"u\c{c}l\"u}
\affiliation{Department of Physics, Izmir Institute of Technology, IZTECH,
  TR35430, Izmir, Turkey}

\date{\today}

\begin{abstract}
Using many-body configuration interaction techniques we show that
Wigner crystallization occurs at the zigzag edges of graphene at
surprisingly high electronic densities up to $0.8$
$\mbox{nm}^{-1}$. In contrast with one-dimensional electron gas, the
flat-band structure of the edge states makes the system interaction
dominated, facilitating the electronic localization. The resulting Wigner
crystal manifests itself in pair-correlation functions, and evolves
smoothly as the edge electron density is lowered. We also show that
the crystallization affects the magnetization of the edges. While the
edges are fully polarized when the system is charge neutral (i.e. high
density), above the critical density, the spin-spin correlations
between neighboring electrons go through a smooth transition from
antiferromagnetic to magnetic coupling as the electronic density is
lowered.
\end{abstract}
\maketitle


Wigner crystallization, i.e. localization of electrons induced by
electron-electron interactions\cite{Wigner34}, remains a key issue in
strongly interacting systems. In an electron gas, as the electronic
density is reduced, the Coulomb repulsion energy overcomes the kinetic
energy and the electrons become localized at their classical
positions. The two limits, high density Fermi liquid and low density
Wigner crystal, are well understood. However, the crossover in between
is a complex many-body problem which was previously investigated for
various electron gas systems in various dimensions both theoretically
\cite{Tanatar+89,Filinov+01,Attaccalite+03,Waintal06,GGU+06,Ghosal+07,GGU+08,Mat04,Casula06,Casula08,Guclu+09,Mehta+13}
and experimentally\cite{KravSarachik04,SAY+06,AYdP+02,ASY+05,DB08,Brun+14}. 
In particular, it
is expected that Wigner crystallization has important implications on
transport properties of two-dimensional\cite{KravSarachik04} and
one-dimensional\cite{Mat04,Guclu+09,AYdP+02,ASY+05,DB08,Brun+14} systems.


For graphene\cite{Wallace47,NGP+09,Novoselov+Geim+04,Zhang+Tan+05}, the
investigation of Wigner crystallization remains limited
\cite{zhang+07,cote+08,wang+scarola12,guerrero+14} partially due to
the fact that for massless Dirac electrons with linear dispersion (as
opposed to quadratic dispersion of free electron gas), the interaction
strength does not depend on electronic density\cite{guclu+book14}. It
is however possible to induce a mass term, for instance, through
application of external magnetic field, for which the Wigner crystal
regime was studied within meanfield theory \cite{zhang+07,cote+08}, or
through size quantization \cite{guerrero+14}. Another situation where
Wigner crystallization in graphene may occur is when zigzag edges are
present as suggested in Ref.\onlinecite{wang+scarola12}. However, as
far as we know, there is no detailed analysis of the many-body problem
of Wigner crystal transition at graphene edges. Indeed, zigzag edges
give rise to a band of half-filled degenerate states near the Fermi
level without the need for an external magnetic field. Electrons
populating these edge states constitute a particularly interesting
many-body system since their relative kinetic energy is close to zero,
thus the properties are dominated by Coulomb interactions.  So far,
most of the previous literature on interaction
effects due to edge states in various graphene systems focused on
magnetic properties
\cite{guclu+book14,Wakabayashi+98,Son+Nat+06,Yazyev+08,Wunsch+08,Jung+09,Wimmer+08,Yazyev+PRB+11,Guclu+13,Ezawa+07,FRP+07,Wang+Meng+08,Guclu+09b,Potasz+12}. In
particular, room-temperature magnetic properties of zigzag edge state
in graphene nanoribbons were recently investigated
experimentally\cite{Magda+14}.  However, for the design of
carbon-based next-generation devices such as
nanoribbons\cite{Wakabayashi+98,Son+Nat+06,Yazyev+08,Wunsch+08,Jung+09,Wimmer+08,Yazyev+PRB+11,Guclu+13}
and quantum
dots\cite{Ezawa+07,FRP+07,Wang+Meng+08,Guclu+09b,Potasz+12}, an
in-depth understanding of Wigner crystallization at graphene edges is
necessary and a focused investigation of the liquid to crystal
crossover is lacking.


In this work, we use a combination of tight-binding method and
configuration interaction technique on a two-dimensional honeycomb
lattice to show that strong Wigner crystallization does indeed occur
at zigzag edges as the electronic density is varied. An analysis of
the pair-correlation functions shows that the critical electronic
density where the localization occur is close to $0.8$
$\mbox{nm}^{-1}$, a value much higher than the critical density for a
one-dimensional (1D) electron gas. Indeed, for the 1D electron gas the
formation of a Wigner crystal was observed using tunneling
spectroscopy into a quantum wire and clear evidence of electron
localization was found at a density of $\rho^{1D} \sim 20$ $\mu
\mbox{m}^{-1}$ \cite{SAY+06} whereas Quantum Monte Carlo calculations
give $\rho^{1D} \sim 15$ $\mu \mbox{m}^{-1}$ \cite{Guclu+09}, both
significantly lower than the critical density at the graphene edges
found in this work. Finally, we investigate ground state magnetization
and spin-spin correlations functions between neighboring electrons to
show that the spin correlations are strongly tied to the formation of 1D
Wigner crystal as a function of electronic density.


In order to model the interaction effects at the zigzag edges, we
start with a graphene ribbon with periodic boundary
condition\cite{Guclu+13}, consisting of $N_{a}=1456$ atoms, with a
length of $L=12.8$ $\mbox{nm}$ and a width of $W=2.9$ $\mbox{nm}$.  This gives $n_s=15$ edge states on
each edge, which are computed using tight-binding technique within
next nearest neighbour approximation of $p_z$ orbitals. The
nearest-neighbour and next-nearest-neighbour hopping elements are
taken to be $t_{nn}=-2.8$ eV and $t_{nnn}=-0.1$ eV \cite{NGP+09}. In
addition, since our main goal is to investigate the Wigner crystal
properties of a single edge, a small electric-field perturbation
perpendicular to the edges was added in order to decouple the 
states belonging to opposite edges.
Next, the fifteen edge states belonging to the upper edge (see Fig.1)
were used to compute two-body scattering matrix elements $\langle p s
\vert V \vert d f \rangle $ in terms of the two-body localized $p_z$
orbital scattering matrix elements $\langle i j \vert V \vert k l
\rangle $. Slater type orbitals\cite{Potasz+12} were used to calculate
the scattering matrix elements. As the overlap between the bulk state
wavefunctions and the edge state wavefunctions is small, electronic
correlations between them is expected to be weak and the configuration
interaction calculations can be performed in the subspace of edge
states\cite{Guclu+13}.  Finally, ground states in subspaces $(N, S_z)$
with different electron number $N$ occupying the edge states and
$z$-component of the total spin $S_z$ are found using diagonalization
of the many-body Hamiltonian given by

\begin{eqnarray}
\nonumber
&H_{MB}&=\sum_{s,\sigma}E_{s}a^\dagger_{s\sigma}a_{s\sigma}
\\&+&\frac{1}{2}\sum_{\substack{s,p,d,f,\\\sigma,\sigma'}} \langle sp\mid V\mid df\rangle
a^\dagger_{s\sigma}a^\dagger_{p\sigma'}a_{d\sigma'}a_{f\sigma}.
\label{eq:CIMBody}
\end{eqnarray}

Here, $E_{s}$ are the kinetic energies in the nearly degenerate shell
of edge states. By comparing the ground state energies of different
$(N,S_z)$ subspaces, it is then possible to deduce the ground state
total spin $S$. In this work, the dimension of the largest matrix we
have diagonalized using Lanczos technique is $2927925 \times 2927925$.

\begin{figure}
\includegraphics[width=\linewidth]{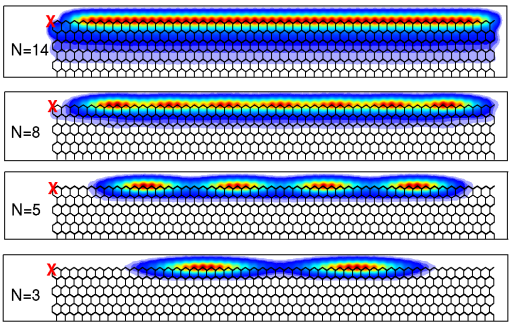}
\caption{Two-dimensional pair-correlation functions for $N=$14, 8, 5,
  and 3 electrons occupying the zigzag graphene edge. The position of
  the fixed electron is indicated by a cross (red online).  At $N=14$
  charge oscillations only at the atomic level are observed. At lower
  $N$ values, $N-1$ peaks arise and become increasingly localized,
  indicating the formation of an one-dimensional Wigner crystal.  }
\label{fig:1} 
\end{figure}
In systems with, e.g. translational or rotational symmetry, electronic
localization can be conveniently investigated through pair-correlation
functions:
\begin{eqnarray}
\nonumber
   P_{\sigma_1 \sigma_2}({\bf r}_1,{\bf r}_2)&=&\langle n_{\sigma_1}({\bf r}_1)
                                                  n_{\sigma_2}({\bf r}_2) \rangle\\
   =\sum_{\sigma_3,...\sigma_N}\int d{\bf r}_3 ... d{\bf r}_N &\times&
      \vert \Psi({\bf r}_1, \sigma_1; ... ;{\bf r}_N, \sigma_N) \vert^2
\label{eq:paco}
\end{eqnarray}
which gives the conditional probability to find an electron with spin
$\sigma_1$ at the position ${\bf r}_1$ provided another electron with spin
$\sigma_2$ is located at ${\bf r}_2$. Figure 1 shows the
pair-correlation functions for different electron numbers $N$ populating
the edge states. The fixed electron has spin up and is located at the
position indicated by a cross. At $N=14$, i.e.  close to charge
neutrality, no charge inhomogeneities (except due to localization over
single atoms) is observed. However, when the density is reduced
oscillations start to appear. At $N=8$, oscillations are weak but seven
peaks (not counting the fixed electron) are observed which is an
indication of Wigner crystallization. At lower densities, localization is
strongly enhanced and the overlap between the electrons is close to zero. 

\begin{figure}
\includegraphics[width=\linewidth]{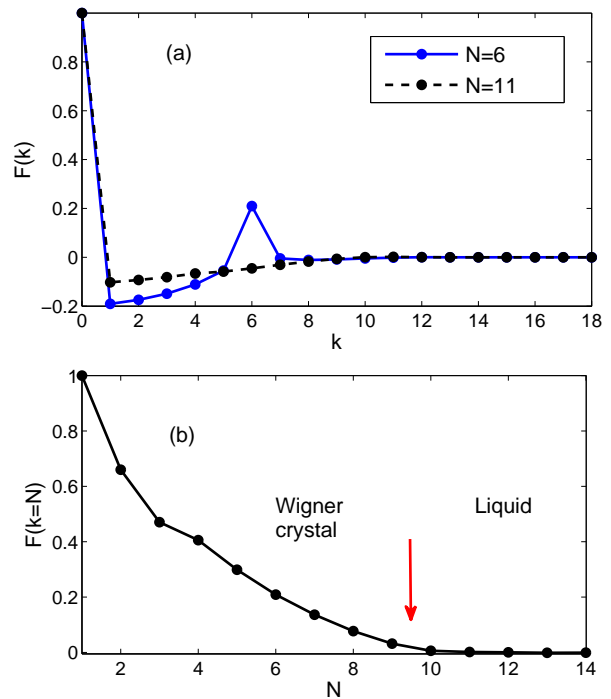}
\caption{(a) Power spectrum as a function of Fourier component $k$ for
  $N=6$ electrons (solid lines, blue online) and $N=11$ electrons
  (dashed lines). For $N=6$, a peak at $k=6$ is observed which is an
  indication of charge localization at classical locations. (b) Power
  spectrum peak height at $k=N$ as a function of $N$. Above $N=10$,
  the peak height is practically zero which indicates a lack of Wigner
  crystallization. The solid-liquid crossover occurs at a
  one-dimensional density of $0.8$ $\mbox{nm}^{-1}$.
}
\label{fig:2} 
\end{figure}
Although the pair-correlation plots are convenient for visualization
of Wigner crystallization, they do not allow to quantify the degree of
localization and to pinpoint the liquid to crystal crossover. This can
be achieved by analyzing the power spectrum\cite{GGU+08}, i.e. the
Fourier transform $F(k)$ of $P_{\sigma \sigma_0}({\bf r},{\bf r}_0)$
in the $x$ direction along the ribbon. In Fig. 2a, we show $F(k)$
for six and eleven electrons. For six electrons, we clearly see a peak
at $k=6$, a signature of electronic localization at their classical
positions\cite{GGU+08}. For $N=11$ however, no localization is observed,
indicating that the electronic density is too high to allow for Wigner
crystallization. In order to pinpoint the electronic density where the
localization occurs, Fig.2b shows the power spectrum peak height
$F(k=N)$ for $N$ up to fourteen. We see that, as the electronic
density is decreased, the peak height decreases, indicating a
transition toward liquid state. Above $N=10$, no localization is
observed. In particular, the system is in a liquid state in the
vicinity of charge neutrality, i.e. $N=15$. 
The crossover value corresponds to
a one-dimensional density of $0.8$ $\mbox{nm}^{-1}$.  This value is
strikingly higher than the critical density for 1D electron
gas for which experimental observations \cite{SAY+06} and theoretical
calculations \cite{Guclu+09} give $n^{1D} \sim 15-20$ $\mu \mbox{m}^{-1}$.

\begin{figure}
\includegraphics[width=\linewidth]{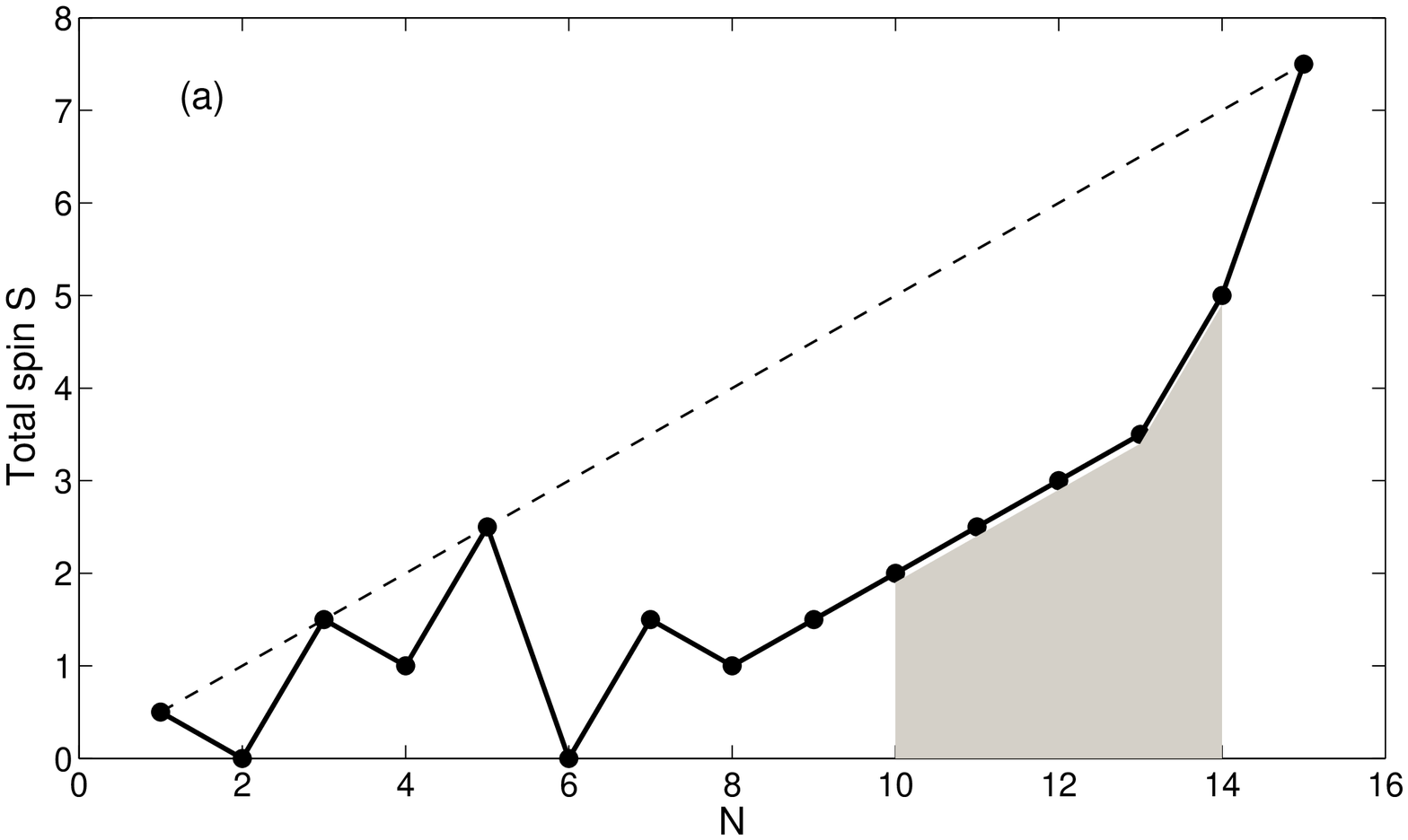}
\includegraphics[width=\linewidth]{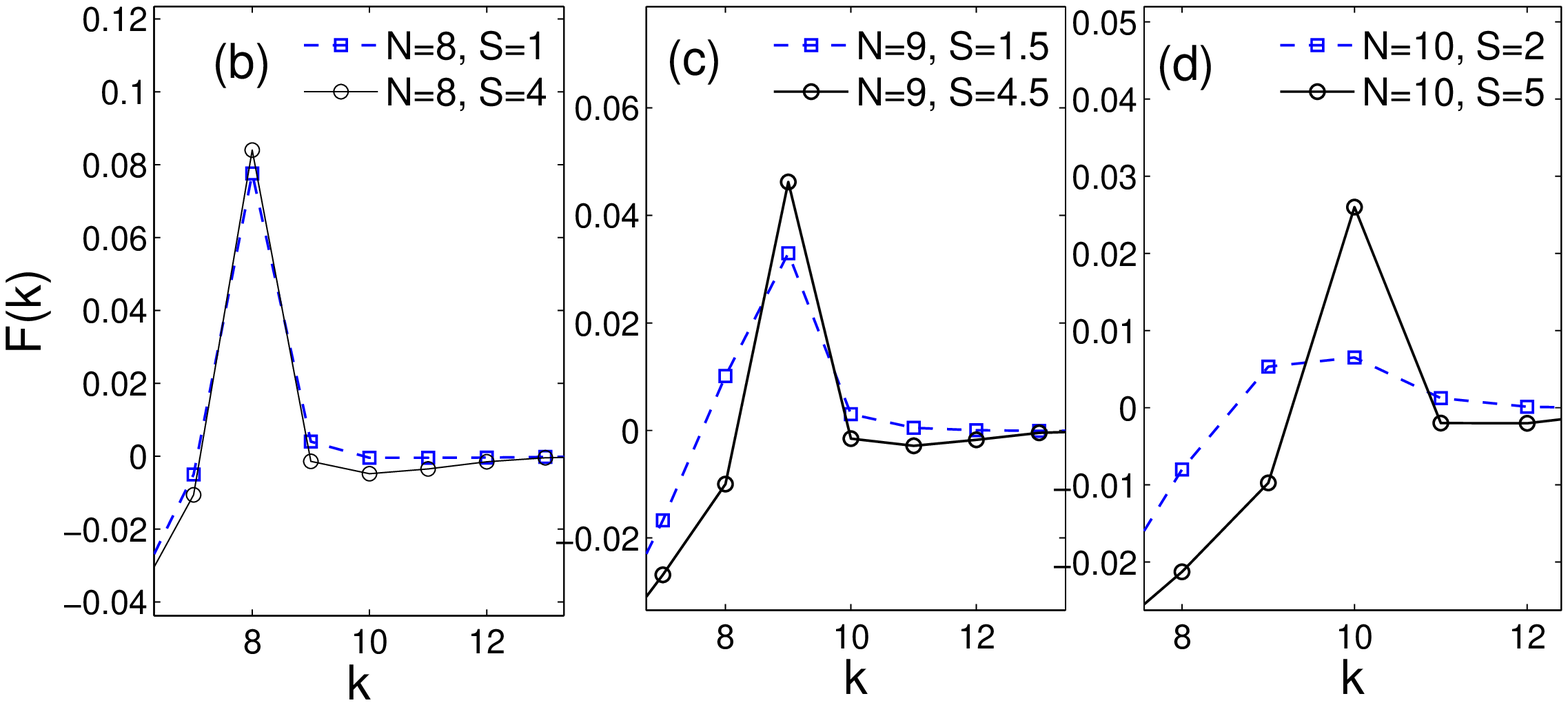}
\caption{(a) Ground state total spin $S$ as a function of number of edge
  electrons $N$. Dashed line shows the maximum possible total spin and
  the shaded area indicates the uncertainty in the total spin due to
  computational limitations. At $N=15$ the system is charge neutral
  and the edges are fully polarized. Away from charge neutrality
  a reduction in magnetization occurs. (b)-(d) Spin dependence of the power
  spectra for $N=$ 8, 9, and 10.   }
\label{fig:3} 
\end{figure}
We now analyze the ground state magnetic properties as a function of
electronic density. Figure 3a shows the ground state total spin $S$ as
a function of number of edge electrons $N$. It is well established
that, in agreement with Lieb's theorem\cite{Lieb+89}, charge neutral
system give rise to ferromagnetic edges. In our case, this means that
for $N=15$, the total spin is $S_{max}=15/2$. However, away from
charge neutrality, correlation effects are expected to strongly affect
the magnetization\cite{Guclu+13}. In Fig.3a, the dashed line shows the
maximum possible polarization. The shaded area indicates an
uncertainty in $S$ due to computational limitations, since it becomes
exponentially more difficult to diagonalize matrices for small values
of $S_z$ at large $N$. Thus, the solid line in this area represents an
upper limit to $S$. However, the uncertainty does not affect our
estimation of the critical density of Wigner crystallization since the
crystallization is already very weak at these $N$
values. Nevertheless, a clear reduction in magnetization, which is
consistent with but more pronounced than in previous calculations for
smaller system sizes\cite{Wunsch+08}, is observed. In Fig.3b-d, we
also investigate the spin dependence of the power spectra for $N=$ 8,
9, and 10. For the fully polarized state, $S=N/2$, the power spectrum
peak height is found to be always higher than the depolarized ground
state, indicating stronger localization. However, the difference
becomes negligible below $N=9$ which is another indication that the
system enters the Wigner crystal regime\cite{GGU+08}.

\begin{figure}
\includegraphics[width=\linewidth]{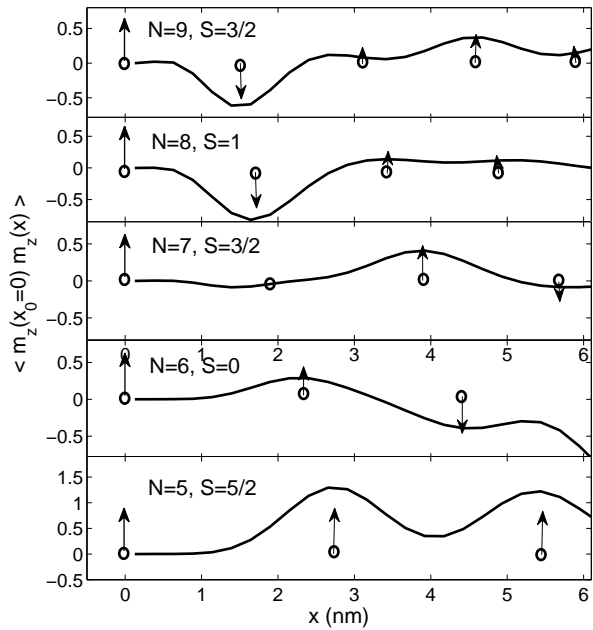}
\caption{Ground state spin-spin correlation function along the
  graphene edge for $N=$ 9, 8, 7, 6 , and 5. The small circles with
  arrows represent the classical position of the localized electrons
  and their effective spin relative to the fixed electron at $x=0$. As
  the density is lowered, the magnetic correlation between the nearest
  neighbors switches from antiferromagnetic coupling to ferromagnetic
  coupling. The vertical scale is kept the same in all the panels.}
\label{fig:4} 
\end{figure}
In order to investigate further the connection between the Wigner
crystallization and the ground magnetization, in Fig.4 we plot the
ground state spin-spin correlation functions $\langle m_z(x_0)m_z(x)
\rangle$ along the edge atoms, where $m_z=n_\uparrow-n_\downarrow$.
The small circles with arrows represent the classical position of the
localized electrons and their effective spin relative to the fixed
electron at $x=0$. For $N=9$, Wigner localization has already started
but it is weak, with a ground state total spin of $S=3/2$. The
spin-spin correlation function at the nearest neighbors is negative,
indicating antiferromagnetic coupling.  For $N=8$, the spin-spin
correlation function does not change significantly compared to the
$N=9$ case.  However, as the electronic density is decreased further,
the average distance between the electrons increases faster, and
the magnetic correlations are affected accordingly. As a result, for $N=7$
and $S=3/2$, the magnetic correlations between the nearest electrons
drops significantly, and become ferromagnetic for $N=6$ and
$S=0$. This ferromagnetic coupling between the nearest electrons is
further enhanced for $N=5$ and $S=5/2$. These results show that the
magnetization of the edges is closely tied to the evolution of the
Wigner crystallization.

To conclude, we have shown that a one-dimensional Wigner
crystallization occurs at the zigzag edges of graphene. An analysis of
pair-correlation functions through configuration interaction
calculations indicates that the crossover from the Fermi liquid to
Wigner solid occurs near a strikingly high critical density of $0.8$
$\mbox{nm}^{-1}$, as compared to the critical density $n^{1D} \sim
15-20$ $\mu \mbox{m}^{-1}$ for the one-dimensional electron gas.
While the spin of the ground state of the charge neutral system is
fully polarized, we observe magnetic depolarization and oscillations
as the liquid-solid crossover occurs. By analyzing the spin-spin
correlations between the neighboring electrons, we have shown that the
magnetic oscillations are accompanied by a transition from
antiferromagnetic to ferromagnetic coupling between the localized
electrons. Localization effects can be observed for instance using
tunneling spectroscopy measurements as was done for a one-dimensional
electron gas\cite{ASY+05}.  Clearly, for the design of carbon-based
spintronic devices, Wigner crystallization must be taken into account
for the full understanding of charge and spin transports. Finally, we
note that although we have considered an ideal edge without any
structural imperfections, inhomogeneities are expected to amplify and
not wash out the liquid to crystal transition\cite{GGU+06}.  Thus, in
more realistic graphene structures Wigner crystallization should be
even more robust, strongly affecting both the transport and spin
properties. Identification of combined effects of imperfections and
interaction induced localization requires further investigations.

{\it Acknowledgment}. This work was supported by The Scientific and
Technological Research Council of Turkey (TUBITAK) under the 1001
Grant Project Number 114F331 and by Bilim Akademisi - The Science
Academy, Turkey under the BAGEP program. The author thanks Pawel
Hawrylak, Harold U. Baranger and Nejat Bulut for valuable discussions.




\end{document}